\documentclass[a4paper,11pt,article,oneside]{memoir}
\usepackage{ecis2015}
\usepackage{amsmath}
\lstdefinelanguage{XML}
{
  morestring=[b]",
  morestring=[s]{>}{<},
  morecomment=[s]{<?}{?>},
  morekeywords={xmlns,version,type,ma-id}
}

\DeclareAcronym{scn}{
      short = SCN ,
      long  = Supply Chain Network ,
      tag = abbrev
    }
\DeclareAcronym{sc}{
      short = SC ,
      long  = Supply Chain ,
         short-plural = s ,
      long-plural = s,
      tag = abbrev }
\DeclareAcronym{kpi}{
      short = KPI ,
      long  = Key Performance Indicator ,
      tag = abbrev
    }
\DeclareAcronym{e2e}{
      short = E2E ,
      long  = End-to-End ,
      tag = abbrev
    }
\DeclareAcronym{cq}{
      short = CQ ,
      long  = Competency Question ,
       short-plural = s ,
      long-plural = s,
      tag = abbrev
    }
\DeclareAcronym{kg}{
      short = KG ,
      long  = Knowlegde-Graph ,
       short-plural = s ,
      long-plural = s,
      tag = abbrev
    }
\DeclareAcronym{oem}{
      short = OEM ,
      long  = Original Equipment Manufacturer,
       short-plural = s ,
      long-plural = s,
      tag = abbrev
    }

    \DeclareAcronym{SENS}{
      short = SENS ,
      long  = SENS,
      short-plural = s ,
      long-plural = s,
      tag = abbrev
    }
 \DeclareAcronym{SCOR}{
      short = SCOR ,
      long  = SCOR,
      short-plural = s ,
      long-plural = s,
      tag = abbrev
    }   
\DeclareAcronym{SENS-GEN}{
      short = SENS-GEN ,
      long  = SENS-GEN,
      short-plural = s ,
      long-plural = s,
      tag = abbrev
    }
    
    \DeclareAcronym{scf}{
      short = SCF ,
      long  = Supply Chain Formation,
      short-plural = s ,
      long-plural = s,
      tag = abbrev
    }
    
\maintitle{SENS: Semantic Synthetic Benchmarking Model for integrated supply chain simulation and analysis} 
\shorttitle{SENS: Semantic Synthetic Benchmarking Model} 
\category{Research Paper} 

\authors{Nour Ramzy, Infineon Technologies AG, Neubiberg, Germany,  nour.ramzy@infineon.com \par
Sören Auer, TIB Leibniz Information Centre for Science and Technology, Hannover, Germany, auer@tib.eu \par

Hans Ehm, Infineon Technologies AG, Neubiberg, Germany, hans.ehm@infineon.com \par

Javad Chamanara, L3S Research Center, Leibniz University of Hannover, Hannover, Germany,\newline
chamanara@l3s.de
} 

\begin{document}
\setlength{\parskip}{0pt}
\begin{abstract}\noindent
\ac{sc} modeling is essential to understand and influence \ac{sc} behavior, especially for increasingly globalized and complex \acp{sc}. 
Existing models address various \ac{sc} notions, e.g., processes, tiers and production, in an isolated manner limiting enriched analysis granted by integrated information systems. 
Moreover, the scarcity of real-world data prevents the benchmarking of the overall \ac{sc} performance in different circumstances, especially wrt. resilience during disruption. 
We present SENS, an ontology-based \ac{kg} equipped with SPARQL implementations of KPIs to incorporate an end-to-end perspective of the~\ac{sc} including standardized SCOR processes and metrics.
Further, we propose SENS-GEN, a highly configurable data generator that leverages SENS to create synthetic semantic \ac{sc} data under multiple scenario configurations for comprehensive analysis and benchmarking applications.
The evaluation shows that the significantly improved simulation and analysis capabilities, enabled by SENS, facilitate grasping, controlling and ultimately enhancing \ac{sc} behavior and increasing resilience in disruptive scenarios.
\end{abstract}

\begin{keywords}
 Ontology , Supply Chain Modeling, Synthetic Data , Benchmarking 
\end{keywords}


\nocite{lin2010semantic}
\nocite{huan2004review}
\nocite{xiao2009optimization}
\nocite{suherman2017network}
\nocite{zdravkovic2011approach}
\nocite{kirikova2012joint}
\nocite{lu2013ontology}

\chapter{Introduction}
 Our increasingly global economy results in a high interconnection between~\acfp{sc}~\citep{soundararajan2021multinational}.
Consequently,~\acp{sc} are evolving from being a chain of businesses with one-to-one relationships to becoming a network of multiple businesses that provide products and services to customers \citep{lambert2000issues}. 
Thus, analyzing the behavior of~\ac{sc} is essential as it does not only affect one organization but a highly complex, dispersed and connected network. 
Namely,~\ac{sc} modeling and benchmarking enable proactive monitoring of processes across the network~\citep{winkelmann2009conceptual}. 

For instance, the \ac{e2e} \ac{scn} model provides an overall perspective of the \ac{sc} partners as well as the flow of products, services, and materials which conveys \ac{sc} structural coherence and resilience. 
The SCOR model provides standard definitions of operational processes and \ac{kpi} to enable \ac{sc} standardization and benchmarking. 
Existing \ac{sc} models tackle core aspects but still in an isolated manner, hence, limiting integrated \ac{sc} behavioral analysis.
Furthermore, the scarcity of integrated empirical data from \ac{sc} members limits the study of the overall behavior. Firms do not disclose their connections to keep a competitive advantage or simply because there are not enough associated incentives or rewards \citep{brintrup2015supply}. Also, logs or data from one company are not enough to validate the end-end \ac{sc} models. 

In this paper, we investigate how semantic models aid the benchmarking of \acp{sc} for better simulation and integrated analysis. 
Our main contribution is SENS, a semantic model that leverages ontologies, \acp{kg} and the SPARQL query language to provide an overall perspective of an \ac{e2e} \ac{scn}, standardized SCOR processes and performance indicators. 
SENS comprises of \ac{sc} partners and the relations between them, representing the flow of materials and goods.
Moreover, based on the production and inventory capacity model included, we provide a SPARQL-based demand fulfillment algorithm that mimics how a \ac{sc} operates to achieve its ultimate goal of meeting end-customers' order requests. 
Additionally, we propose SENS-GEN, a highly configurable synthetic data generator that, using input parameters and SENS model, produces an exemplary instance of an \ac{scn}.
SENS-GEN enables the generation of \ac{sc} data for various industries, e.g., automotive and dairy, determined by the topology and properties of the instantiated output \ac{kg}. 
As a result, companies can rely on SENS and SENS-GEN to generate data for various simulated \acp{sc} in order to apply analysis methods and make informed decisions faster and more reliable. 
Ultimately, we deem that better simulation and analysis, as put forward by SENS, will contribute to standardizing and benchmarking \acp{sc}, thus mastering more complex \ac{sc} scenarios, increasing the resilience of supply networks and ultimately facilitating digitalization.
 

The remainder of the paper is structured as follows: in Section~\ref{chap:literature}, we give an overview of the literature on existing \ac{sc} models namely SCOR, \ac{e2e} \ac{scn} and semantic models while examining the core \ac{sc} aspects they tackle. 
In Section~\ref{chap:model}, we present SENS, our \ac{sc} model that incorporates \ac{sc} core aspects in an integrated manner. 
We propose SENS-GEN, a configurable data generator that leverages the SENS ontology to create a particular synthetic realization of an \ac{scn} i.e., SENS \ac{kg} in Section~\ref{chap:generator}.
In Section~\ref{chap:evaluation}, we evaluate SENS as an integrated \ac{sc} model that enables the simulation of \ac{sc} behavior in experimental contexts for comprehensive performance analysis.
Finally, we conclude by presenting the limitations, implications and outlook of our contribution.

\chapter{Background and Motivation}
\label{chap:literature}
Modeling aids the understanding and monitoring of structural and operational aspects within \acp{sc}. 
In this section, we review the \ac{sc} concepts incorporated by SCOR, \ac{e2e} \ac{scn} and semantic models as they address essential aspects such as  standardization, coherence, interoperability and information integration.

\section{Supply Chain Models}

\paragraph{\ac{SCOR} Model.}
To evaluate \ac{sc} performance and continuously improve, \ac{sc} standardization offers a mutual understanding of concepts and processes, consequently enabling benchmarking and comparison of performance.  
The classic \ac{SCOR} model, introduced by APICS\footnote{\url{https://www.ascm.org/}} in 1997,  provides a common terminology to define \ac{sc} standardized  activities and performances~\citep{SCC2010scor}. 

The \ac{SCOR} model covers all customer interactions (order entry through paid invoice); we refer to this as \textbf{(C1)}. 
Additionally, it spans all physical material transactions \textbf{(C2)} and all market interactions (from the understanding of aggregate demand to the fulfillment of each order) \textbf{(C3)}.
Also, the \ac{SCOR} model contains standard descriptions of the~\ac{sc} processes e.g., \textit{Source}, \textit{Plan}, \textit{Make}, \textit{Deliver}, \textit{Enable} and \textit{Return}, \textbf{(C4)}.
Furthermore, the \ac{SCOR} model organizes~\ac{sc} performance metrics, i.e., \ac{kpi}, into a hierarchical structure \textbf{(C5)} to determine and compare the performance of \ac{sc} on various levels e.g., top strategies, tactical configurations and operational processes \citep{irfan2008scor}. 
In addition, \ac{SCOR} describes best-in-class management practices \textbf{(C6)} and maps software products that enable best practices \textbf{(C7)}.  
In order to gain an overall perspective of \ac{sc} operational performance and structural coherence, \ac{e2e} \ac{scn} models are fundamental. 

\begin{figure}[h]
\centering
\includegraphics[width=\linewidth, trim=0cm 0cm 0cm 0.05cm, clip=true, height=5cm]{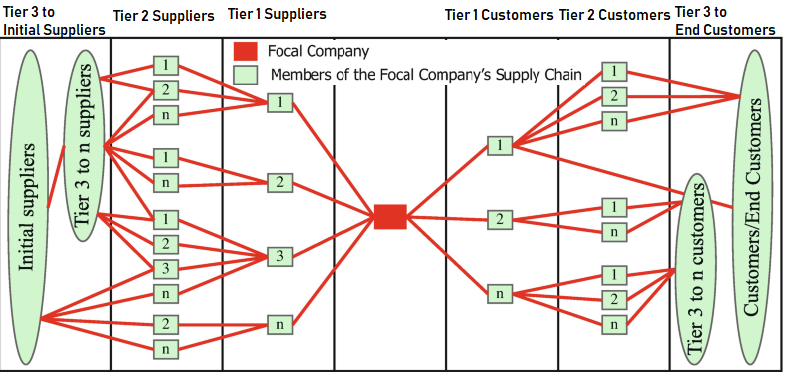}
\caption{Supply chain network structure by \citep{lambert2000issues}.}
\label{fig:scnetwork}
\end{figure}

\paragraph{End-to-End Network Models.} 
An \ac{scn} is a network representation of the physical nodes of a \ac{sc} and how they relate to one another \citep{golan2020trends}. 
The \ac{e2e} model provides an overall perspective of the \ac{sc} nodes topology that starts at the procurement of raw materials and ends at the delivery of finished goods to the end customers. 
The literature review by \citep{bier2018formation} highlights key \ac{sc} aspects in an \ac{e2e} \ac{scn} model.
The authors identify that an~\ac{scn} consists of a representation of vertices i.e., nodes acting as \ac{sc} partners, \textbf{(C8)}. 
~\ac{sc} partners are connected with edges \textbf{(C9)} modeling product, demand flow and contractual relations as shown in \autoref{fig:scnetwork}.
Nodes are organized in tiers, nodes in the same tier supply goods and services for the following tiers.
 
An~\ac{scn} model considers various materials used to manufacture the end product \textbf{(C10)}. The authors describe that the focal company, i.e., \ac{oem}, distinguishes between supply and demand flows, i.e., \textbf{(C11)}. 
Partners in the~\ac{scn} can be facilities, companies, or warehouses. Nevertheless, the competition in the future will be~\ac{sc} vs.~\ac{sc} where each node participates in one or more~\acp{sc} \textbf{(C12)} while sharing and competing with other nodes over suppliers and customers \citep{chain2001supply}.  
Due to the diversity, dispersion and complexity within an~\ac{scn}, interoperability is challenging. 
However, relying on semantic models enables information exchange and allows partners to reach full and agile information integration. 

\paragraph{Semantic Models.}
Semantic models have been developed as an attempt to represent the complexity of the \ac{sc} domain, e.g.,~\citep{ye2008ontology} developed Onto-SCM to provide shared terminologies for ~\ac{sc} concepts and relations. 
The literature review by~\citep{grubic2010supply} lists existing~\ac{sc} ontologies to model the \ac{sc}'s key concepts. 
The authors identify that a semantic model includes the strategic, tactical and operational views of the~\ac{sc} \textbf{(C13)}. 
Besides, an~\ac{sc} ontology covers an organizational extent i.e., internal or external \textbf{(C14)}. 
The model incorporates an industry sector \textbf{(C15)}, has a purpose \textbf{(C16)} and supports~\ac{sc} applications \textbf{(C17)}. 

\section{Gap Analysis}
We examine the literature reviews by \citep{delipinar2016using}, \citep{bier2018formation} and \citep{grubic2010supply} for existing SCOR, \ac{e2e},  semantic models respectively. We identify the gap between the artifacts in the studied models and the previously listed \ac{sc} aspects \textbf{(C1-17)}. 

\paragraph{Gap Analysis for~\ac{sc} Models.}
We note that existing \ac{sc} \ac{SCOR} models do not include management practices and software products \textbf{(C6, C7)} as they are considered sensitive information to keep a competitive advantage \citep{delipinar2016using}.  
Moreover, \citep{bier2018formation} create a comparison framework of~\ac{sc} \ac{e2e} network models and conclude that the academic literature does not contain studies that address the topology of~\ac{scn} \textbf{(C8, C9)} together with detailed insights on structural information \textbf{(C10, C11)}.
Additionally, emergent \ac{scn} topology literature include \ac{sc} nodes operations independently and not as part of one or many~\ac{sc} \textbf{(C12)} \citep{brintrup2018supply}. 

Furthermore, all the existing~\ac{sc} ontologies cover the strategic level of granularity,  yet none supports tactical and operational levels \textbf{(C13)} \citep{grubic2010supply}.  For \textbf{(C14)}, existing models are limited to the inter-business network scope. 

\paragraph{Gap Analysis for Hybrid~\ac{sc} Models.}
In an attempt to fulfill the shortcomings of existing models, we study hybrid models that combine \ac{SCOR}, \ac{e2e},  semantic \ac{sc} models pair-wise.
\autoref{tab:gapanalysis} lists the literature for \ac{sc} hybrid models and identifies gaps with respect to the concepts \textbf{(C1-17)}. 
We highlight, in gray, the \ac{sc} concepts that are not covered by existing \ac{sc} models discussed in the previous section.
\begin{table} [h]
\includegraphics[width=\linewidth]{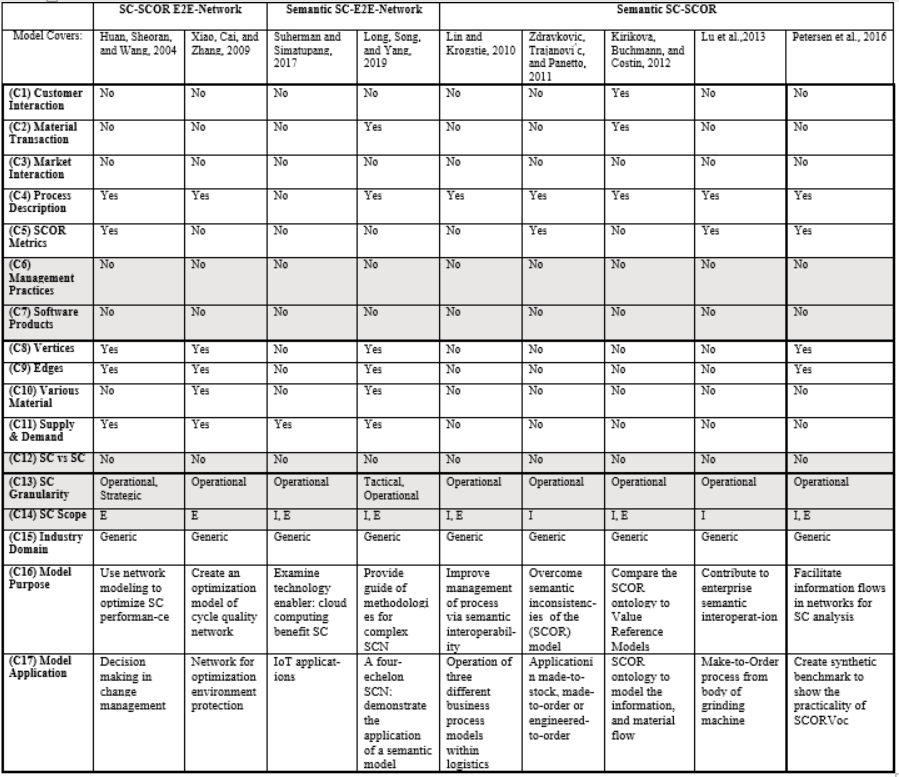}
  \caption{Gap analysis of existing \ac{sc} models and covered \ac{sc} concepts. E: External, I: Internal.}
\label{tab:gapanalysis}
\end{table}
In the gap analysis process, we consider different models as follows:
\begin{enumerate}
  \item We examine models that combine \ac{SCOR} and \ac{e2e}~\ac{scn} and the corresponding \ac{sc} concepts i.e., \textbf{(C1-7), (C8-12)}. We observe that \textbf{(C4), (C8), (C9)} and \textbf{(C11)} are incorporated by existing approaches in the literature. Namely, the model by~\citep{xiao2009optimization} includes \ac{SCOR} metrics \textbf{(C5)} and various raw materials \textbf{(C10)}. However, existing models do not cover the following \ac{SCOR} notions: customer interactions, material transactions, market interactions, management practices and software products.

  \item We study models incorporating \ac{e2e}   \textbf{(C8-12)} and  semantic \textbf{(C13-17)} concepts. 
  Only the work of \citep{long2019semantic} offers a semantic model addressing all aspects of an \ac{e2e} \ac{scn} model \textbf{(C8-11)}, except \textbf{(C12)}. 
  Also, the authors include the tactical and operational granularity levels \textbf{(C13)}. Both proposed works cover an internal and external \ac{sc} scope \textbf{(C14)}. 
  
  \item We consider semantic \ac{SCOR} models \textbf{(C1-7)} and \textbf{(C13-19)}. We mention only the attempts at semantic modeling of SCOR-based \acp{sc} revised in latest works. All models listed in \autoref{tab:gapanalysis} address \ac{SCOR} \ac{sc} aspects \textbf{(C4)}, \textbf{(C5)}, however, we note that \textbf{(C1), (C2), (C3), (C6)} and \textbf{(C7)} are not satisfied. While none of the models is industry-specific \textbf{(C15)}, they include only the operational granularity of a \ac{sc} \textbf{(C13)} though, they provide a purpose and an application, respectively \textbf{(C16)} and \textbf{(C17)}.
\end{enumerate}


\section{Motivation and Contribution} 
Integrated modeling of the \ac{sc} enables enriched behavioral analysis and benchmarking, as it incorporates various \ac{sc} concepts, e.g., materials, metrics and processes, while giving a holistic perspective of the \ac{sc} structure, flows and partners. 
Existing \ac{SCOR}, \ac{e2e} and semantic models alongside corresponding hybrid models are limited as they convey essential \ac{sc} aspects in an isolated manner.
Also, the scarcity of empirical data from multiple \ac{sc} partners hinders the analysis of the overall impact of supply network partners on each other. Available data and logs from one company are not enough for \ac{e2e} benchmarks.  
Therefore, we propose SENS-GEN, a highly configurable data generator that relies on the SENS integrated semantic model to generate a synthetic \ac{sc} instance for standardization and benchmarking of an \ac{e2e} \ac{scn}. 
\raggedbottom

\chapter{SENS: Integrated Semantic Supply Chain Model} \label{chap:model}
We present SENS, an integrated semantic \ac{sc} model that incorporates an end-to-end perspective of the~\ac{sc} including standardized SCOR processes and metrics \acp{sc}.  

\section{SENS Ontology Model} 
\begin{figure}
  \includegraphics[width=\linewidth, trim=0cm 0cm 0cm 0.5cm, clip=true]{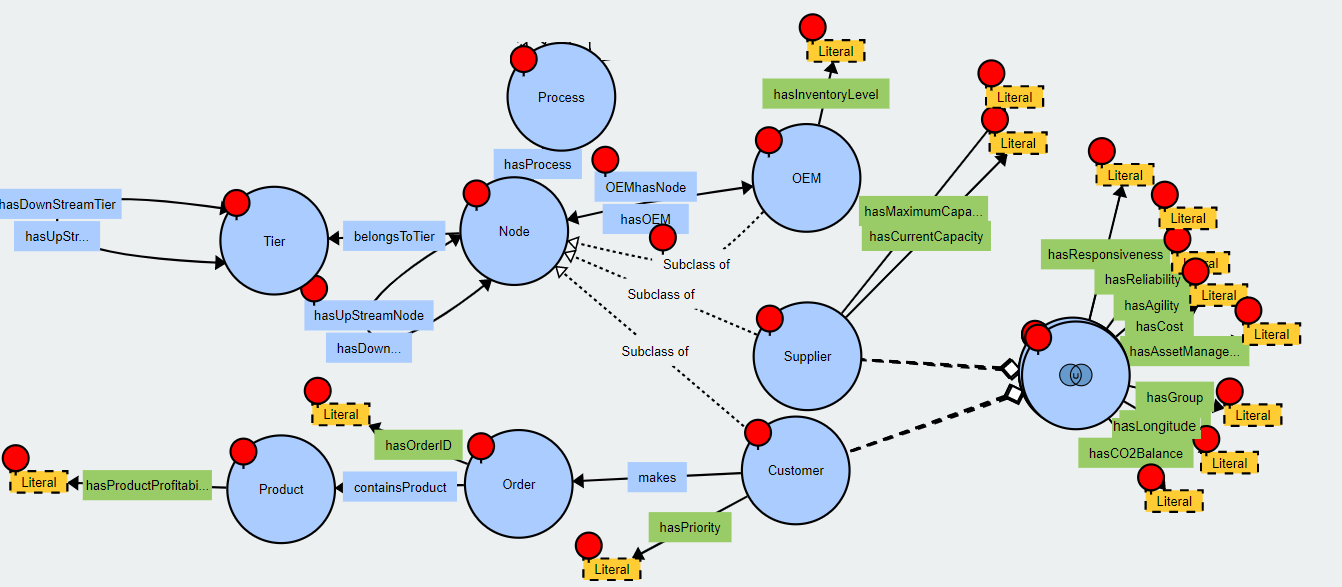} 
  \caption{Depiction of the core concepts of the SENS ontology modeling End-to-End and SCOR  supply chain concepts.}
   \label{fig:scontology}
\end{figure}

The core of SENS Ontology depicted in \autoref{fig:scontology} is nodes representing \ac{sc} partners.
We model each partner as instance of the class \textit{Node}, i.e., \textit{Supplier}, \textit{Customer} or \textit{\ac{oem}}. 
\ac{sc} nodes are organized in tiers, so we model this information using RDF triples of the form \textit{Node belongsToTier Tier}.
Accordingly, we distinguish between \textit{SupplierTier} and \textit{CustomerTier}. 

The supply side is organized so that the raw material suppliers belong to the highest supplier tier, which is the most upstream tier, i.e., \textit{SupplierTierN} \citep{brintrup2018supply}.
Supplier nodes in low tiers are connected to suppliers in upstream tiers using the property \textit{hasUpStreamNode} while on the customer side, end customers belong to the most downstream tier,  i.e., \textit{CustomerTierN}. 
Similarly, customer nodes in the low customer tier are connected to customers at downstream tiers with the property \textit{hasDownStreamNode}. 
The links between nodes model contractual relations, organizing the flow of demand, materials and products between \ac{sc} partners.
Likewise, \textit{SupplierTier}s are connected with \textit{hasUpStreamTier} while \textit{CustomerTier} with \textit{hasDownStreamTier}.

The Original Equipment Manufacturer (\ac{oem}) is the focal node responsible for assembling the product or getting it ready for distribution by delivering it to a warehouse or a wholesaler, followed by various distribution centers to the end-customer. 
The \ac{oem} is directly linked to the suppliers in \textit{SupplierTier1} via the property \textit{hasOEM} and \textit{CustomerTier1} via \textit{OEMhasNode}

Also, we model node operations with RDF triple statements of the form \textit{Node hasProcess Process} and the class \textit{Process} has as subclasses the SCOR processes: \textit{Source, Plan, Make, Deliver, Enable and Return}. 
Consequently, for each node, we model the SCOR \ac{kpi} \textit{hasResponsiveness, hasReliability, hasCost, hasAgility, hasAssetManagementEfficiency} to evaluate the operational behavior of this node based on the SCOR metrics standard. 
Furthermore, each node is described by data properties that either depict its performance e.g., \textit{hasCO2Balance} or its characteristics e.g., \textit{hasLocation}. We resolve node locations using geo-coordinates represented with the properties \textit{hasLongitude, hasLatitude}.  

\section{Supply Chain Demand Fulfillment}
The goal of an \ac{scn} is to fulfill end-customers' demand relying on production and inventory capacities. SENS models supply and demand and a SPARQL-based demand fulfillment algorithm to simulate \ac{sc} production planning and scheduling.
\paragraph{Supply Chain Demand.}
We model the demand as orders of products via triples of the following form:
\textit{Node  makes Order}, 
\textit{Order hasProduct  Product}, 
\textit{Order hasDeliveryTime  xsd:dateTime} and \textit{Order hasQuantity  xsd:integer}. 
Moreover, customer orders are fulfilled depending on their priority modeled by \textit{Node hasPriority xsd:integer}. Customer relationship management determines a customer's priority based on various factors, e.g., customer revenue, contract type. 

\paragraph{Supply Chain Capacity and Production.}
\ac{sc} nodes produce and stock products in order to fulfill the demand.
We rely on RDF-star, a framework to model in a compact way statements about statements \citep{rdfstar}.  RFD-star is widely implemented by tools such as GraphDB and Virtuoso;  reification \citep{Patel-Schneider:14:RS} is a viable alternative.
The following list of triples models capacity and production of nodes in the \ac{scn}: 

\begin{itemize}
\item \textit{Node manufactures Product}: defines what products are manufactured by this node e.g., \textit{\ac{oem} manufactures Car}. \textit{<<Product needsProduct Product>> needsQuantity xsd:integer} models the intermediate products needed to manufacture the final product.  For instance, \textit{<<Car needsProduct Wheel>> needsQuantity '4'} and \textit{<<Wheel needsProduct Rubber>> needsQuantity 10m}.  
\item \textit{Node hasTransportMode xsd:string}: \ac{sc} nodes rely on one or more shipment modes e.g., air cargo, maritime to transport products. 

\item \textit{Node hasGroup xsd:integer}: in order to reduce purchasing prices and benefit from the supreme performance, suppliers capable of supplying the same products, i.e.,  belong to the same group, are exchangeable \citep{hofstetter2019multi}.

\item \textit{Node hasCapacity Capacity}: defines the availability of labour and resources to make a product by a node.  The capacity is detailed by  \textit{Capacity hasProduct Product},  \textit{Capacity hasCost xsd:integer}, \textit{Capacity hasQuantity xsd:integer} and \textit{Capacity hasTimeStamp xsd:dateTime}. 

\item \textit{Node hasSaturation xsd:integer}: is the bottleneck defining the maximum capacity to manufacture at any time. 

\item \textit{Node hasInventory Inventory}: models the node keeping stock of products describing the inventory using triples of the following form: \textit{Inventory hasProduct Product; hasCost xsd:integer; hasQuantity xsd:integer; hasTimeStamp xsd:dateTime}.

\item \textit{Node hasDeliveryTime xsd:integer}: indicates the time for a node to deliver to the customer after finishing production \citep{xiao2016two}. 
\end{itemize}

\paragraph{Demand Fulfillment.}
\acp{sc} follow a customer order-based strategy to determine its production scheduling \citep{borgstrom2011supply}. 
We present a SPARQL-based demand fulfillment algorithm relying on backward scheduling, i.e., starting from the delivery time of an order and planning backward for its fulfillment. 
The \textbf{input} is incoming orders containing a standard product with constant repetitive demand.  The \textbf{output} of this algorithm is a supply plan specific for each order modeled by \textit{Order hasSupplyPlan SupplyPlan}. 
This plan is a scheduled capacity allocation for products among production facilities as well as the needed parts among suppliers as shown by the following triple representation: \textit{<<SupplyPlan needsNode Node>> getsProduct Product; hasTimeStamp xsd:dateTime; hasQuantity xsd:integer; hasUnitPrice xsd:double}.

We determine the following base assumptions about the model: 
\begin{itemize}
    \item Nodes have a standard delivery time. When the node capacity is lower than the saturation limit,  i.e., the node is operating far from the bottleneck, orders are fulfilled and delivered in constant time \citep{cannella2018capacity}. 
    
    \item The supplier selection process is based on respective capacities while suppliers' choice can potentially consider other factors, e.g., price, quality of service or CO2 balance \citep{setak2012supplier}.
    
    \item  The demand fulfillment is a recursive cascading problem, e.g., nodes in \textit{TierN} receive orders from nodes in \textit{TierN+1}.
    Then, the fulfillment either relies on the available inventory or production capacities. On the supply side,  nodes in \textit{TierN} decompose the product to the intermediate products supplied by nodes in \textit{TierN-1}, whereas on the customer side, the same finished ordered products flow between nodes. 
  
    \item \ac{sc} planners determine the frequency of execution of the demand fulfillment algorithm. 
\end{itemize}
In  this  sense,  we  consider  the  relationships  between  three tiers  of the \ac{sc} (SupplierTier1, \ac{oem} and CustomerTier1). 
The incoming demand to the \ac{oem} is the orders by customers in CustomerTier1 and is the aggregation of the incoming demand flow starting from the end-customer. 
 
The following steps, executed at time \textit{t}, outline the demand fulfillment algorithm. For conciseness, we show exemplary queries while we provide the detailed code and SPARQL queries in our accompanying technical report and GitHub repository \citep{NRamzy2021}. 
\begin{enumerate} 
    \item \autoref{alg_orders}: 
    At \textit{t}:  Get orders by customer priority from CustomerTier1  where \textit{O rdf:type Order}, \textit{O hasProduct P},  \textit{O hasDeliveryTime $DT(O)$}. The \ac{oem} has delivery time modeled by \textit{\ac{oem} hasDeliveryTime $LT(O)$} where $DT(O) - LT(O) = t$ 
    \begin{lstlisting}[caption= Get Orders by customer priority,label=alg_orders]
SELECT * WHERE {?o hasDeliveryTime ?dt. ?o hasQuantity ?q. ?o hasProduct ?p. ?cus makes ?o. ?cus hasPriority?prio. ?oem hasDeliveryTime ?lt. FILTER (?dt-lt=t)} ORDER BY DESC ?prio
\end{lstlisting}
   
    \item 
    If \ac{oem} inventory at \textit{t hasQuantity Q(I)} suffices to fulfill the order quantity i.e., \textit{O hasQuantity $Q(O)$} and $Q(I)>=Q(O)$, then the order is fulfilled, a supply plan generated and the \ac{oem} inventory updated: $Q(I)= Q(I)-Q(O)$. Otherwise, we proceed with production in step 3.  
     
    \item Place a production order for the remaining $Q(I)- Q(O)$, if the \ac{oem} capacity at \textit{t} is smaller than its saturation.
    \begin{enumerate}
     \item  \autoref{getcomponents}:
     Get all intermediate products and quantities to manufacturer P. \begin{lstlisting} [caption= Get all intermediate products for Product P,label=getcomponents ]
SELECT * WHERE { << P needsProduct ?comp >> needsQuantity ?quant.}
\end{lstlisting}
    \item \autoref{getsupplier}: 
    Choose a supplier in SupplierTier1 with capacity for intermediate products smaller than the bottleneck at $t_0$ with $t_0= t-LT(S)$, where \textit{Supplier hasDeliveryTime $LT(S)$}. This means that the supplier has the capacity to produce the intermediate products at $t_0$ to reach the \ac{oem} at \textit{t} to manufacture and fulfill the order at its delivery time  $DT(O)$. If suppliers are chosen for all intermediate products, then the order is fulfilled and a supply plan generated. Otherwise, the order is not fulfilled.  
 
\begin{lstlisting} [caption= Get Supplier capacity for intermediate product at time $t_0$,label=getsupplier]
SELECT * WHERE {?s hasOEM OEM1. ?s hasCapacity ?cap. ?cap hasProduct ?p. 
?cap hasQuantity ?q. ?cap hasTimeStamp ?t0. ?s hasSaturation ?sat. ?s hasDeliveryTime ?lt. FILTER  (?sat>= ?q +  tofullfil) &&  (t - ?lt= ?t0).}
\end{lstlisting}
    \end{enumerate}
\end{enumerate}

 \raggedbottom

\chapter{SENS-GEN: Synthetic Supply Chain Knowledge Graph Generator} \label{chap:generator}
This section presents SENS-GEN, a highly configurable data generator that relies on the SENS model to create a specific synthetic instance of an \ac{scn}, incorporating \ac{sc} concepts in an integrated manner. 

\section{SENS-GEN Parametrization} 
SENS-GEN receives input parameters to instantiate  SENS ontology, i.e., SENS \ac{kg}, that determines the topology and the performance of the \ac{scn}.
Namely, the topology depends on the industry sector as it signifies the complexity of the products (the steps needed to manufacture), the variability and the number of customers and suppliers.
In fact, the topology is defined by the \textit{Supplier\_Tier, Node\_Supplier\_Tier, Customer\_Tier, Node\_Customer\_Tier} parameters in \autoref{tab:inputparameters}.

The \ac{kg} describes the behavior of the \ac{scn} through the values assigned to the nodes' data properties e.g., \textit{hasReliability, hasCO2Balance}. Namely, the capacity and inventory of the nodes allow the simulation of the demand fulfillment and evaluate the performance of this particular \ac{sc} realization.  
The parameters assigned per node can be randomly generated from the range of values given, e.g., [1-5], or manually defined per node as an input. 
For conciseness, we show only the supplier side generation in Algorithm \autoref{alg:cap} (cf. the technical report \citep{NRamzy2021} for the detailed code).

\begin{algorithm}
\begin{algorithmic}

\caption{SENS knowledge-graph generation algorithm}\label{alg:cap}
\For{\texttt{($n=1$; $n<=Supplier\_Tier$; $n++$)}} \Comment{Create tiers and nodes}
         \State Create SupplierTier\textit{(n)} 
         \For{\texttt{($m=1$; $m<=Node\_Supplier\_Tier[n]$; $m++$)}}
         \State Create SupplierNode\textit{(m.n)}
         \State Add SupplierNode\textit{(m.n)}, \textit{:hasGroup}, Random(1, Supplier\_Group\_Tier[n])
         \For{\textit{Property P of SupplierNode(m.n)}}\Comment{e.g., \textit{:hasCO2Balance}}
         \State Add SupplierNode\textit{(m.n)}, p , Random(min\_val, max\_val)  
         \State Generate  saturation capacity, initial capacity and inventory
   \EndFor
   \EndFor
    \EndFor

\end{algorithmic}    
\end{algorithm}

\begin{table}[h]
\small
  \begin{tabular} {|p{.32\textwidth}|p{.32\textwidth}|p{.13\textwidth}|p{.13\textwidth}|}
    \hline
    \bigstrut[t]          \textbf{Parameter }\newline \textbf{Triple Representation}   & \textbf{Explanation} & \textbf{Automotive Industry} & \textbf{Dairy Industry} \\
    \hline
    \bigstrut[t] 
     Supplier\_Tier & \textbf{\ac{sc}} depth, manufacturing steps 
     & 3 & 1 \\ \hline
    Customer\_Tier & \textbf{\ac{sc}}  distribution and sales interactions  (OEM to end customer)  & 3 & 2 \\ \hline    
    
    Node\_Supplier\_Tier  & \textbf{\ac{sc}}  width, the suppliers providing  materials for manufacturing & <2, 3, 5> & <3>   \\ \hline
    
    Node\_Customer\_Tier   &\textbf{\ac{sc}} customer availability& <2, 2, 4> & <2, 3>  \\ \hline

    Supplier\_Group\_Tier\newline \textit{Supplier hasGroup xsd:integer} & \textbf{Supplier} exchangeability to provide same products per tier & <1, 2, 4> &  <1> \\ \hline
    
    Node\_Priority range \newline \textit{Node hasPriority xsd:integer} & \textbf{Customer} relationship management to prioritize customers & [1-3] &  [1-3] \\ \hline
    
    Node\_Capacity\_Saturation\newline \textit{Node hasSaturation xsd:integer} & \textbf{Node} maximum  capacity to manufacture & [1-3] million unit &  [0.5-1] million unit\\ \hline

    Node\_Delivery\_Time \newline \textit{Node hasDeliveryTime xsd:integer}& \textbf{Node} time to deliver from node to node in following tier & [1-7] days & [1-3] days\\ \hline 

 Node\_Initial\_Inventory \newline {\textit{Node hasInventory Inventory}}& \textbf{Node} inventory at t=0 & [10-50] thousand unit & [5-10] thousand unit \\\hline
 
 	Node\_Initial\_Capacity \newline \textit{Node hasCapacity Capacity}& \textbf{Node} capacity at t=0 & 1 thousand unit & 1 thousand unit \\\hline
 	
Data Property range  \newline \textit{Node (hasResponsiveness, hasReliability, hasCost, hasAgilty, hasAssetMangmentEfficeny) xsd:integer} & \textbf{SCOR} KPIs. \cite{petersen2016scorvoc} explain how to calculate  level 1 SCOR \ac{kpi} from lower level metrics for SCOR processes  &  [0-100] \%&  [0-100] \% \\ \hline

Data Property range \newline \textit{Node hasCO2Balance xsd:integer}  & \textbf{\ac{sc}} environmental performance &  [30-45] Tg &  [30-45] Tg  \\ \hline

Data Property range  \newline \textit{Node hasLongitude xsd:integer} \textit{Node hasLatitude xsd:integer}  & \textbf{\ac{sc}} globalization (geographically dispersed
network of nodes)  &   Long/Lat: [0-180/ 0-90] &   Long/Lat: [90-180/ 45-90]  \\ \hline

Customer\_Demand\_Frequency  \newline \textit{Customer makes Order} & \textbf{\ac{sc}} constant demand frequency & 2 & 10  \\ \hline
Product type and quantity per order \newline \textit{Order hasProduct Product} \newline \textit{Order hasQuantity xsd:integer}& \textbf{\ac{sc}} orders variability and size  & 1: 100 thousand unit & 1: 5000   \\ \hline
 
  \end{tabular}
  \caption{SENS-GEN parametrization and exemplary parameters for automotive and dairy industry.}
  \label{tab:inputparameters}
\end{table}
\raggedbottom

\section{Generated Showcase Examples}
\begin{figure}[h]
  \includegraphics[width=\linewidth, trim=0cm 0cm 0cm 0cm, clip=true]{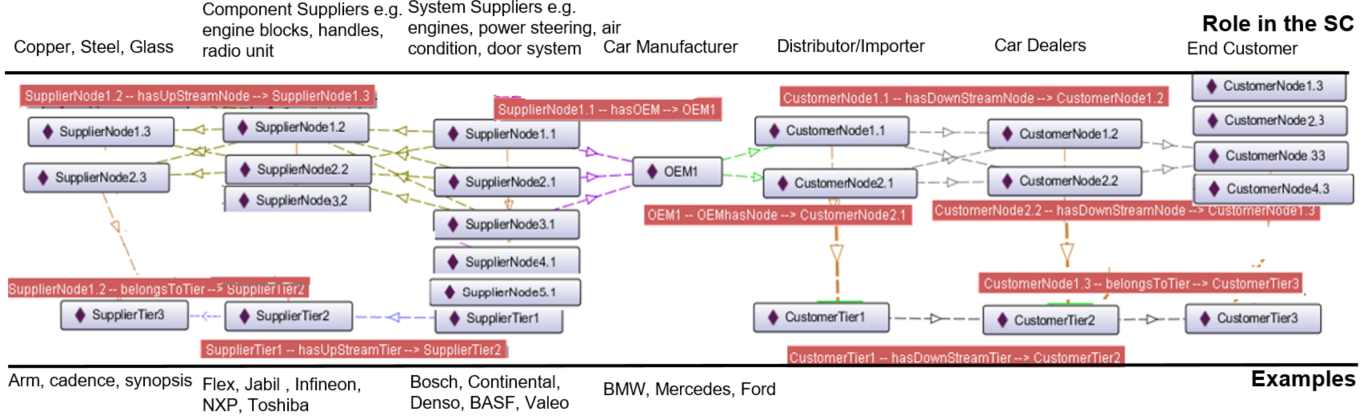} 
  \caption{Automotive industry SENS \ac{kg} example with three supplier tiers raw material, component and system suppliers.}
   \label{fig:e2ekg}
\end{figure}

\begin{figure}[h]
 \includegraphics[width=\linewidth]{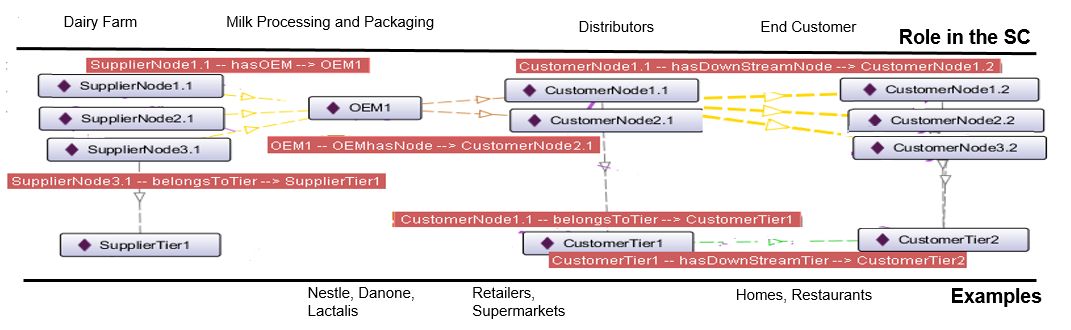} 
  \caption{Dairy industry SENS \ac{kg} example with one supplier tier, i.e., the dairy farms and one end-customers tier.}
   \label{fig:e2ekg2}
\end{figure}

We present two examples of \ac{scn}s from the automotive and dairy industries. \autoref{tab:inputparameters} shows the parametrization of the model and the variation of topology and properties based on the industry.
In \autoref{fig:e2ekg}, we provide an example of a \ac{scn} in the automotive industry.
We choose three supplier tiers, i.e., raw material, component and system suppliers.
The dairy \ac{scn} example in \autoref{fig:e2ekg2} 
consists of one supplier tier, i.e., the dairy farms that are directly linked to the \ac{oem}.
At the \ac{oem}, products are processed and packaged to be sent to retailers  \texttt{CustomerTier1} then end-customers \texttt{CustomerTier2} e.g., homes, restaurants. 

There exist multiple \ac{kpi}s to assess \ac{sc} behavior, yet we focus on the SCOR KPIs as they enable a  standardized performance evaluation and benchmarking.  We set for the \ac{SCOR} \ac{kpi}, a range of [0-100]\% as explained by \citep{petersen2016scorvoc}.
The CO2 balance varies according to policies of countries where nodes are located as well as \ac{oem} environmental strategies but range between 30-45 Teragram (Tg) \citep{thoma2013greenhouse}. 
Since the dairy products are easily perishable, dairy \acp{sc} are not dispersed.
The range for longitude, latitude and inventory is smaller and the delivery time is shorter than in the automotive industry. 
However, in the dairy industry, customer orders are more frequent but include smaller product quantities. 

\raggedbottom
\chapter{Evaluation}
\label{chap:evaluation}

\section{SENS Evaluation}
We perform a two-fold evaluation. First, we prove that SENS is a semantic \ac{sc} model that integrates core aspects of \ac{sc} and deals with shortcomings caused by isolated models.  
Then, we provide an empirical performance analysis of the generated automotive \ac{scn} example introduced in Section \ref{chap:generator} and show behavioral changes under experimental conditions.

\paragraph{SENS Model Validation.}
We validate that SENS is an integrated model by analyzing SENS  coverage of  \ac{sc} concepts \textbf{(C1-17)} incorporated by SCOR, \ac{e2e} and semantic \ac{sc} models, listed in our literature assessment.
In \autoref{tab:evaluationtable}, we show the executed SPARQL queries and sample results from the automotive SENS \ac{kg}. 
\begin{table}[h]   
\small
  \begin{tabular}{|p{.25\textwidth}|p{0.35\textwidth}|p{.30\textwidth}|} 
    \hline
    \bigstrut[t]         &  \textbf{SPARQL Query:} \newline SELECT  \text{*} 
WHERE & \textbf{Example Output Triples}  \\
    \hline
     \textbf{(C1) Customer Interaction} & 
     ?customer makes ?order. \newline?customer hasDownStream ?c & Node3.2 makes OrderJZHu5
       Node3.2  hasDownStream Node3.3\\\hline 
       
      \textbf{(C2) Material Transaction / (C10) Various Materials} & 	 
      <<Product needsProduct ?p>> needsQuantity ?q & 
     <<ProductA needsProduct Product1>> needsQuantity 1 
  \\\hline
     \textbf{(C4) Process Description}& 
   ?node hasProcess ?process. 
   &  Node3.2 hasProcess ProcessA. ProcessA rdf:type Make   \\\hline
      
       \textbf{(C5) SCOR Metrics}   &
        ?node hasResponsiveness ?r.
      &  Node3.2 hasResponsiveness '24' 
     \\\hline
    
       \textbf{(C8)Vertices / (C9) Edges} & 
    ?node a Node
        ?node ?prop ?node2. 
       &  Node3.2 rdf:type Node
       Node3.2 hasDownStreamNode Node3.3
       \\\hline
   
   \end{tabular}
  \begin{tabular}{|p{.25\textwidth}|p{.677\textwidth}|} 
   \textbf{(C3) Market Interaction / (C11) Supply and Demand}& Algorithm described in Section ~\ref{chap:model} detailed by \citet{NRamzy2021} \\ \hline
     
      \textbf{(C12) SC vs SC}& Supplier \textbf{exchangeability} is modeled  by \textit{Supplier hasGroup xsd:integer}.  Nodes share and compete over suppliers and customers. \\ 
      \hline
      
      \textbf{(C13) SC Granularity}& \textbf{Operational}: SENS-SC spans SCOR operational processes e.g.  \textit{Source, Plan} and the supply plans address operational planning. \textbf{Tactical, Strategic}:  Describing the performance via data properties e.g. \textit{hasCO2Balance} enable analysis on different aggregation levels. 
       \\\hline

      \textbf{(C14) SC Scope} & SENS-SC models \textbf{Internal} node processes and \textbf{External} interactions by modeling the flow of supply and demand.   \\\hline
    \textbf{(C15) Industry Domain} &     Model \textbf{parametrization} in \autoref{chap:generator} to tailor the \ac{kg} to any  industry.
    \\\hline
    
    \textbf{(C16) Model Purpose}& Provide a topology of \ac{scn} with detailed and \textbf{standardized} operational SCOR processes and relying on semantics for \textbf{interoperability}.\\\hline
    
    \textbf{(C17) Model Application}& \ac{sc} \textbf{behavior analysis} in empirical scenarios as shown in the following section. \\\hline

    \end{tabular}

  \caption{SENS as an integrated semantic  model covering \ac{sc} core aspects.}
\label{tab:evaluationtable}
\end{table}
We note that the proposed SENS ontology and \ac{kg} enable us to model and retrieve \ac{sc} aspects \textbf{(C1-17)} except \textbf{(C6, C7)}.
However, existing research in the domain implies that management practices and software products are hard to assess and thus not commonly represented in \ac{sc} models.
We can conclude that SENS integrates \ac{sc} aspects covered by SCOR, \ac{e2e} and semantic \ac{sc} model. 

\paragraph{SENS Knowledge Graph Behavior Analysis.}
This section shows the benchmarking and integrated analysis in experimental contexts, enabled by SENS.  
\subparagraph{Setup:} We use the automotive SENS \ac{kg} in \autoref{fig:e2ekg} generated via the parameters in \autoref{tab:inputparameters}. We run the demand fulfillment algorithm for 178 \textit{t} (days), i.e., half a year.

\subparagraph{Metrics:} The following metrics are a sample of the SPARQL-based performance indicators to benchmark the performance of a semantic \ac{e2e} SCOR \ac{sc}.  
\textbf{Order Fulfillment} in \autoref{orderfull} evaluates how many orders the \ac{sc} fulfills and generates corresponding supply plans.  
This metric quantifies the \ac{sc} ability to achieve its goal of satisfying end customers' demand. 
Also, operating close to the saturation capacity entails longer delivery times and straining production labor and machinery. Thus,    
\textbf{Node Utilization} in \autoref{utilization} measures the extent to which a node employs its installed productive capacity after executing the demand fulfillment algorithm.  
\textbf{Average SCOR KPI} in \autoref{scorkpi} is an example to calculate the average responsiveness of the \ac{sc} nodes. This metric allows the estimation of the speed at which a \ac{sc} provides products to the customer.
    \begin{lstlisting} [caption= Order Fulfillment,label=orderfull]
SELECT  ?order (SUM(IF(REGEX(str(?x),"True"), 1, 0)) AS ?fulfill)
(SUM(IF(REGEX(str(?x),"False"), 1, 0)) AS ?notfulfill) WHERE { ?order isFulfilled  ?x. }
\end{lstlisting}

\begin{lstlisting} [caption= Node Utilization,label=utilization]
SELECT 100*?quant/?max WHERE { ?supplier hasSaturation ?max. ?supplier hasCapacity ?cap. ?cap hasQuantity ?quant.  ?cap hasTimeStamp 178.}
\end{lstlisting}

\begin{lstlisting} [caption= Average SCOR KPI ,label=scorkpi]
SELECT AVG(?res) AS ?Responsiveness WHERE { ?supplier hasResponsiveness ?res. }
\end{lstlisting}

\subparagraph{Parameter variation:} We measure the performance of the \ac{sc} under various experimental scenarios by changing the input parameters Customer\_Demand\_Frequency,  Node\_Capacity\_Saturation. 

 \begin{figure}[h]
  \includegraphics[width=\linewidth, trim=0cm 0cm 0cm 0cm, height=4cm, clip=true]{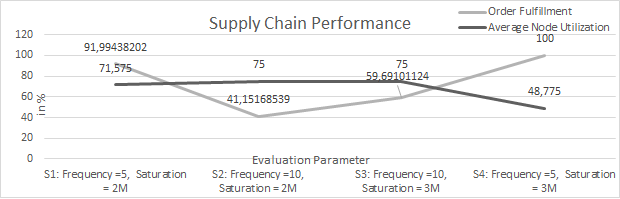} 
  \caption{SENS knowledge graph performance evaluation with parameter variation.}
   \label{fig:evaluation}
\end{figure}
The graph in \autoref{fig:evaluation} shows that the order fulfillment metric drops when the demand frequency doubles
(on the x-axis S1-S2), which is a potential scenario during, e.g., the holidays season. 
Recovering with increasing saturation capacity can help the \ac{sc} perform better as we can see in the graph the surge in order fulfillment from S2 to S3 where Node\_Capacity\_Saturation increased from 2M to 3M.
Moreover, we note that the node utilization is reduced when the Node\_Capacity\_Saturation increases. This result is logical as the nodes are not operating close to their production saturation. 
This is a required setup as it guarantees operational stability and constant delivery time. The average responsiveness is 85\% and does not change with parameter variations.

\section{Discussion}
Including the SCOR model into SENS provides a standardized representation of \ac{sc} processes and \ac{kpi}. 
The \ac{e2e} perspective brings an overall view of the \ac{sc} partners and their relations and flow of supply and demand.
Integrating these models using semantic artifacts facilitates the benchmarking of the overall \ac{sc} behavior. 
\paragraph{Limitations and Next Steps.}
We assume the nodes' characteristics to be constant throughout the simulation.
As a result, the SENS parametrization is rigid to some extent, while real-life scenarios might impose some fuzziness. 
Thus, we propose as a next step to include a degradation function representing deterioration in behavior. 
For instance, the model should include delay functions for transit lead times or a variation of the SCOR \acp{kpi} in different operational conditions, e.g., to reduce responsiveness under high utilization. 
In addition, we generate parameter values randomly or via user input. 
As a next step, we will implement an interactive interface where the user can tailor the values for each node individually to fine-tune the parameter space. 

Also, we can extend SENS to optimize for additional node performance characteristics such as carbon footprint, service level and price.
This will enable extending the implemented supplier choice to include multi-factor-based decision making as explained by \citep{setak2012supplier}. 

The evaluation of SENS and the presented examples cover the basic flows at this stage.
Thus, we will further assess SENS and SENS-GEN in light of concrete real-world use cases. 
The goal will be to validate that SENS can cater to the specific characteristics entailed by the complexity of the manufactured product.

\paragraph{Implications.}
SENS is a semantic model, resembling a digital twin, that facilitates information exchange and integration, thus allowing an optimized control in complex \ac{sc} scenarios \citep{barykin2021place}. 
For instance, \citep{ivanov2021digital} elaborate that \ac{sc} digital twins enable integration to discover the link between \ac{sc} disruption and performance deterioration.
The strategic and tactical information integration in the overall \ac{sc} enabled by SENS increases visibility.
This, in turn, may lead to dramatically reducing demand distortion, i.e., the bullwhip effect \citep{blomkvist2020improving}. 

Furthermore, semantic modeling provides a human and machine-understandable representation of the domain. 
Therefore, we see implications of SENS, an ontology-based model, on the reproducibility and re-useability of \ac{sc} models.
Other \ac{sc} modeling research areas, e.g., \ac{scf} and simulation can rely on SENS to ease the extraction of \ac{sc} configurations for \ac{scf} \citep{ameri2019modeling} or to standardize the creation of simulation models as proposed by \citep{ramzy2020first}.




\raggedbottom
\chapter{Conclusion}\label{chap:conclusion}
\ac{sc} modeling is of monumental importance for globalized and complex \ac{scn}. 
There exist several \ac{sc} models, e.g., SCOR, \ac{e2e}, that incorporate  \ac{sc} artifacts, e.g., operations, production scheduling and flow of materials. 
We identified that existing models comprise \ac{sc} core concepts but in an isolated manner, thus hindering integrated \ac{sc} performance analysis.
Moreover, 
the lack of real-world collective \ac{sc} data constrains empirical behavioral analysis and performance benchmarking required for particular circumstances, e.g., resilience simulation under disruption. 

With SENS, we proposed a semantic \ac{sc} model that integrates \ac{sc} concepts. 
SENS leverages a well-defined ontology, SPARQL queries to include SCOR model artifacts, e.g., processes and performance indicators, as well as an end-to-end perspective to model \ac{sc} partners and the flow of goods and materials. Moreover, SENS includes production and inventory capacity models and a SPARQL-based demand fulfillment algorithm. 
Consequently, SENS ensures \ac{sc} standardization, topological and operational coherence and integration.
Additionally, we propose SENS-GEN, a highly configurable data generator that leverages the SENS model to produce exemplary data based on input parameters and create a specific synthetic instance of a \ac{scn}. 
Namely, SENS-GEN generates synthetic data to simulate \ac{sc} behavior in controlled and designed scenarios.
\ac{sc} stakeholders can rely on SENS and SENS-GEN to assess, benchmark and control complex \ac{sc} scenarios, to determine operational strategies and \ac{sc} structure, increase the resilience and ultimately enable digitalization.
\raggedbottom

\section*{Acknowledgment}

This work has received funding from the project CoyPu - Cognitive Economy Intelligence Platform for the Resilience of Economic Ecosystems (grant 01MK21007A) within the program Federal Ministry for Economic Affairs and Climate Action (BMWK) Innovation Competition on Artificial Intelligence.
\printbibliography
\end{document}